          [2001/04/25 1.1 (PWD)]
\documentclass[a4paper,twocolumn]{esapub} % European paper
\usepackage{natbib}

\usepackage{psfig}
\usepackage{epsfig}
\usepackage{graphicx}
\title{First results on the HMXRB Pulsar SAX~J2103.5+4545 with INTEGRAL}
\author{L.Sidoli}
\author{S.Mereghetti}
\affil{IASF Milano, Italy}
\author{S.Larsson}
\affil{Stockholm Observatory, Sweden}
\author{M.Chernyakova}
\affil{ISDC Versoix, Switzerland}
\author{I.Kreykenbohm}
\affil{IAAT/ISDC Versoix, Switzerland}
\author{P.Kretschmar}
\affil{MPE Garching/ISDC Versoix, Switzerland}
\author{A.Paizis}
\affil{IASF Milano/ISDC Versoix, Switzerland}
\author{A.Santangelo}
\author{C.Ferrigno}
\affil{IASF Palermo, Italy}

\begin{document}


\maketitle

\begin{abstract}

We report on the preliminary timing and spectral analysis of the High Mass X-ray Binary Pulsar SAXJ2103.5+4545 as observed with INTEGRAL during the Galactic Plan Scan of the Core Program.
The source shows a hard spectrum, being detected up to 100 keV. 
The timing analysis performed on IBIS/ISGRI data shows evidence for a spin-up with respect to previous observations, performed in 1997 with BeppoSAX.
\end{abstract}

\section{Introduction}

SAX J2103.5+4545 is a transient HMXRB pulsar 
with a $\sim$358~s pulse period discovered with the WFC on-board BeppoSAX during  an  outburst in 1997 (Hulleman et al., 1998).
Its orbital period of 12.68~days has been
 found with the RXTE during the 1999 outburst (Baykal et al., 2000). 
The likely optical counterpart, a Be star with a magnitude
V=14.2, has been recently discovered (Reig \& Mavromatakis, 2003; Reig et al. 2004). 

During the outburst in 1999, Baykal et al. (2002) observed for the first time,
 with RXTE, the
transition from the spin-up phase to the spin-down regime, while the X--ray flux was declining. 
Indeed, the source underwent a spin-up phase during the initial 
part of the outburst (during which the pulse period decreased by $\sim$0.9~s
in 150~days), 
then the flux dropped (and the pulse frequency saturated), and,
as the flux continued to decline, a weak spin-down phase started.
Moreover, a correlation between spin-up rate and X--ray flux was observed (Baykal et al., 2002), suggestive of the formation of an accretion disk during the periastron passage.

A very preliminary spectral analysis of INTEGRAL public observations of the source
region, performed in Dec.2002 during the performance verification phase, 
has been reported by Lutovinov et al. (2003).

Inam et al. (2004) observed a soft spectral 
component (blackbody with a temperature of 1.9~keV) and 
a transient 22.7~s QPO during a XMM-Newton observation 
performed in Jan, 2003.
 
SAX J2103.5+4545 has been observed several times during the Galactic Plane Scan (GPS) 
which INTEGRAL performs every 12 days as part of the Core Program.
We report here the timing and spectral analysis of these observations.

\begin{small}
\begin{figure*}
\centering
\includegraphics[width=13.0cm]{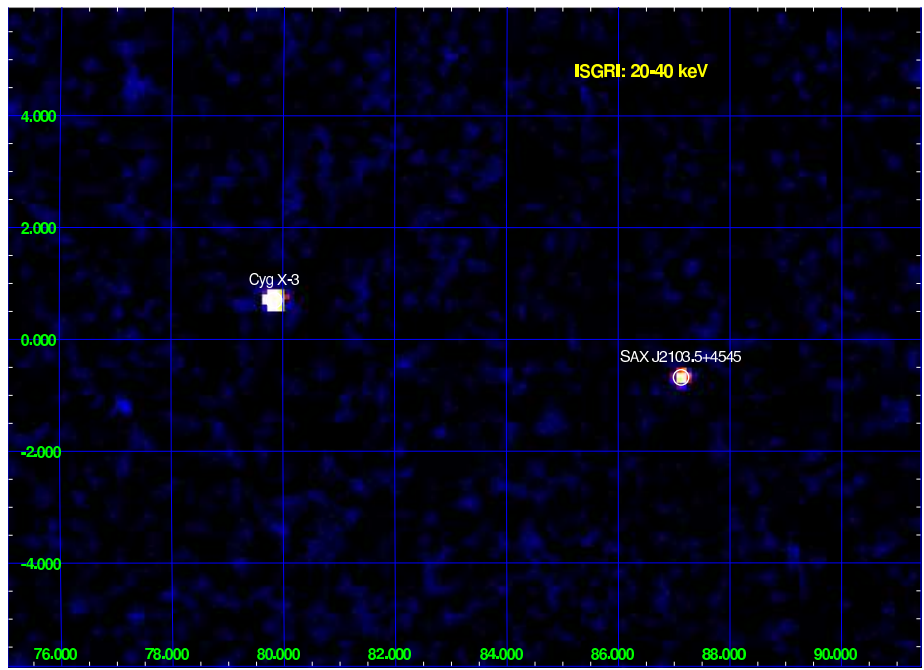}
\vskip -3cm
\caption{IBIS/ISGRI mosaic of the source field in the 20--40~keV energy range in galactic coordinates}
\end{figure*}
\end{small}

An IBIS/ISGRI mosaic of the region of the sky containing SAX J2103.5+4545 is shown in
Fig.~1.
The source long-term lightcurve, as measured with 
RXTE All Sky Monitor (ASM) is
shown in Fig.~2.
The times of the two previous outbursts observed with BeppoSAX (in 1997) 
and RXTE (in 1999) are indicated, as well as the 
epoch of the INTEGRAL observations reported here.

\begin{small}
\begin{figure}
\centering
\includegraphics[width=8.0cm]{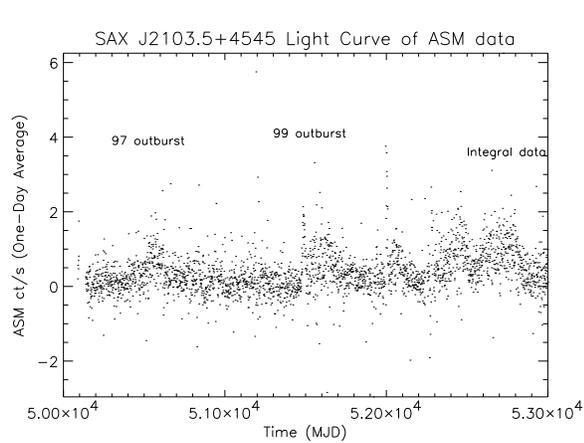}
\caption{RXTE ASM lightcurve of SAX~J2103.5+4545. 
Times of the INTEGRAL observations
have been marked}
\end{figure}
\end{small}

\section{Timing Results}

We have analysed 20 GPS pointings (duration of about 2000~s each) 
covering the region of the sky containing the pulsar.
This sky region is not covered during the Galactic Center Deep Exposure.

In Fig.~3 we show the IBIS/ISGRI lightcurve of SAX~2103.5+4545, where
each point corresponds to the flux from a single pointing.

%%%%%%%%%%%%%%%  FIGURE 3   %%%%%%%%%%%%%%%%%%%%%%%%%%%%%%%%%%%%%%%%
\begin{small}
\begin{figure}
\centering
\includegraphics[width=8.0cm]{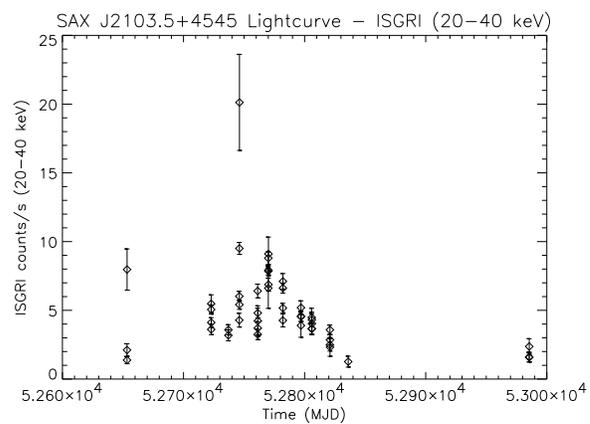}
\caption{SAX~J2103.5+4545 lightcurve with IBIS/ISGRI (20--40~keV). Each point
represents the flux during a single observation 
($\sim$2000~s exposure time)}
\end{figure}
\end{small}
%%%%%%%%%%%%%%%%%%%%%%%%%%%%%%%%%%%%%%%%%%%%%%%%%%%%%%%%%%%%%%%%%%%%

The data have been reduced using OSA3 release of the analysis software.
For each pointing we extracted events with a 
Pixel Illumination Function (pixel fraction illuminated
by the source), PIF, equal to 1.
After  correcting times to the solar system barycenter,
we searched for periodicity around the known pulse period.
We have found a clear peak in the $\chi^2$ distribution 
in 10 observations with IBIS/ISGRI, and only in 3 with JEM-X. 
The estimated pulse periods for each pointing 
are reported in the Table~1, while in Table~2 we report 
the measurements of the pulse period obtained adding together 
two consecutive pointings, and doing the same search for periodicities
on the new data-set. 
The uncertainties on the pulse periods have been estimated
from the Leahy function (Leahy 1987).

 %%----------------------------------------------------------
\vspace{1.5cm}

\begin{table*}[ht!]
\caption{Observations summary and source pulse periods measured  
with IBIS in each pointing}
\vspace{0.5cm}
\label{tab:ppulse}
\begin{center}
\begin{tabular}[c]{llccc}
\hline\noalign{\smallskip}
ID.  &  ISGRI rate      & Start Time &   Pulse Period  &   Pulse Period   \\
  & 20-40 keV (s$^{-1}$)        &  (MJD)       & with ISGRI (s)               & with JEM-X (s)              \\
\noalign{\smallskip\hrule\smallskip}
 1 &       5.06 $\pm{0.34}$ &             52722.88 &
 349 $\pm{7}$  &  334 $\pm{10}$ \\
 2 &      3.58 $\pm{0.38}$ &             52737.07 &
 366 $\pm{10}$ &   $-$  \\
 3 &       6.02 $\pm{0.36}$ &             52746.10 &
 341 $\pm{9}$  &   $-$   \\
 4 &       5.41 $\pm{0.33}$ &            52746.13 &
 364 $\pm{15}$ &   $-$   \\
 5 &       4.81 $\pm{0.35}$ &              52761.29 &
 350 $\pm{10}$ &   $-$   \\
 6 &      8.80 $\pm{0.46}$ &           52770.02 &
 360 $\pm{5}$  &    $-$  \\
 7 &    7.85 $\pm{0.34}$ &             52770.05 &
 357 $\pm{7}$  &   $-$   \\
 8 &    7.92 $\pm{0.35}$ &             52770.08 &
 348 $\pm{6}$  &  368 $\pm{13}$  \\
 9 &      6.61 $\pm{0.36}$ &            52782.09 &
 353 $\pm{8}$  &  351 $\pm{27}$  \\
 10 &      7.10 $\pm{0.57}$ &            52782.12 &
 369 $\pm{14}$ &   $-$   \\
\noalign{\smallskip\hrule\smallskip}
\end{tabular}
\end{center}
\end{table*}

%%%%%%%%% TABLE 2 %%%%%%%%%%%%%%%%%%%%%%%%%%%%%%%%%%%%%%%%%%%%%%%%%%%%%%%%

\begin{table}
  \begin{center}
    \caption{Best determination of the pulse period with IBIS/ISGRI, adding
together two consecutive pointings: the first period have been obtained
searching for periodicities in the new data-set obtained 
adding together the two pointings
n.3 and 4 in Table~1, while the second one considering the sum of the two observations
named n.7 and 8 in Table~1}\vspace{1em}
    \renewcommand{\arraystretch}{1.2}
    \begin{tabular}[h]{lcc}
      \hline
      Start Time       &  Stop Time  &  Period  \\
        MJD            &     MJD     &    s     \\
      \hline
  52746.10 &    52746.15 &  353$\pm{4}$ \\
  52770.05 &    52770.10 &  350$\pm{2}$ \\ 
  \hline \\
      \end{tabular}
    \label{tab:table2}
  \end{center}
\end{table}

%%%%%%%%%%%%%%%%%%%%%%%%%%%%%%%%%%%%%%%%%%%%%%%%%%%%%%%%%%%%%%%%%%%%%%%%%%% 

Our best determination of the pulse period, P, is 350$\pm{2}$~s (ISGRI, 20-40 keV)
indicating a clear  spin up with respect to RXTE and BeppoSAX estimates 
(see Figs.~4 and 5 for the $\chi^2$ distribution and the pulse profile, respectively).

%%%%%%%%%%%%%%%% ---> FIGURE 4  %%%%%%%%%%%%%%%%%%%%%%%%%%%%%%%%%%%%%%%%%%%%%%%%%%%%%%%
\begin{small}
\begin{figure}
\centering
\includegraphics[width=8.0cm]{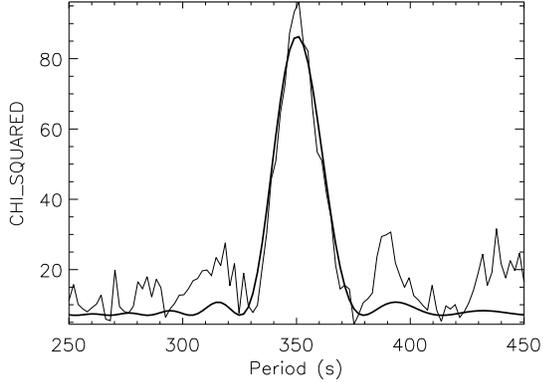}
\caption{$\chi^2$ distribution for our best measure of the pulse period, obtained
summing together two consecutive GPS pointings (see Table~2). The
two lines indicate the data (thin line) 
and the fit with the Leahy function (thick line)}
\end{figure}
\end{small}

%%%%%%%%%%%%%%%%%%%%%%%%%%%%%%%%%%%%%%%%%%%%%%%%%%%%%%%

%%%%%%%%%%%%%%%% ---> FIGURE 5  %%%%%%%%%%%%%%%%%%%%%%%%%%%%%%%%%%%%%%%%%%%%%%%%%%%%%%%
\begin{small}
\begin{figure}
\centering
\includegraphics[width=8.0cm]{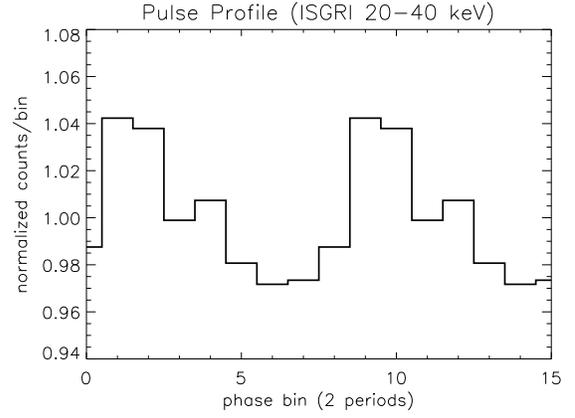}
\caption{Pulse profile for SAX~2103.5+4545  in the energy range 20--40~keV}
\end{figure}
\end{small}

%%%%%%%%%%%%%%%%%%%%%%%%%%%%%%%%%%%%%%%%%%%%%%%%%%%%%%%

A period derivative of about -4$\times$10$^{-7}$~s~s$^{-1}$ 
can be measured with respect to the latest RXTE measurement (P=354.794; Inam et al. 2004) performed in Jan, 2003 indicating an increasing spin-up rate during the last outburst (see Fig.~6). 
We caution that the pulse period history is poorly sampled (Fig.~6), making the
measurement of the period derivative quite uncertain. 
Anyway, our estimate is compatible with the extrapolation at higher X--ray fluxes
($\sim$10$^{-9}$~erg~cm$^{-2}$~s$^{-1}$) of the correlation between
the period derivative  and the X--ray flux, 
measured by Baykal et al. (2002) during the source outburst in 1999
(see their Fig.~7).

%%%%%%%%%%%%%%%% ---> FIGURE 6 %%%%%%%%%%%%%%%%%%%%%%%%%%%%%%%%%%%%%%%%%%%%%%%%%%%%%%%
\begin{small}
\begin{figure}
\centering
\includegraphics[width=8.0cm]{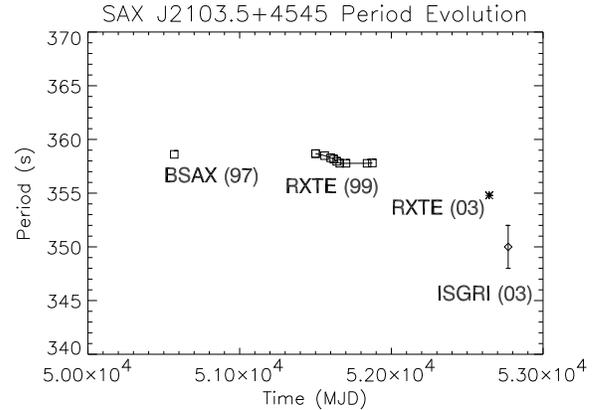}
\caption{SAX~J2103.5+4545 pulse period history; BSAX (97) marks the determination
performed by Hulleman et al. (1998); RXTE (99) is from Baykal et al. (2002);
RXTE (03) has been taken from the recent measurement performed by Inam et al. 2004;
ISGRI (03) is our best measure with IBIS/ISGRI data}
\end{figure}
\end{small}

%%%%%%%%%%%%%%%%%%%%%%%%%%%%%%%%%%%%%%%%%%%%%%%%%%%%%%%

\section{Spectral Results}
 
The results on the spectral analysis from the
three instruments JEM-X, IBIS/ISGRI and SPI, should
be considered very preliminary. 
Indeed, still
large calibration and inter-calibration uncertainties exist.

The overall average source spectrum from 5 to 200 keV, obtained
combining JEM-X together with IBIS/ISGRI and SPI, is quite hard.
We here fitted it with a cut-off powerlaw (including free relative
normalizations between the three instruments).
The resulting 
spectral parameters are a photon index of $\sim$1.1 
and a high energy cutoff of 30 keV (see Fig.~7). 
The 2--100 keV flux is $\sim$1.5$\times$10$^{-9}$~erg~cm$^{-2}$~s$^{-1}$ 
(based on JEM-X and SPI response matrices).

The derived spectral parameters are consistent with   the
WFC/BeppoSAX model (Hulleman et al., 1998). 
%and the bright state observed with RXTE in November, 1999 (Baykal et al. 2002).

%%%%%%%%%%%%%%%% ---> FIGURE 7 SPECTRUM %%%%%%%%%%%%%%%%%%%%%%%%%%%%%%%%%%%%%%%%%%%%%%%%%%%%%%%
\begin{small}
\begin{figure}
\centering
\vskip -0.5cm
\includegraphics[width=5.5cm,angle=-90]{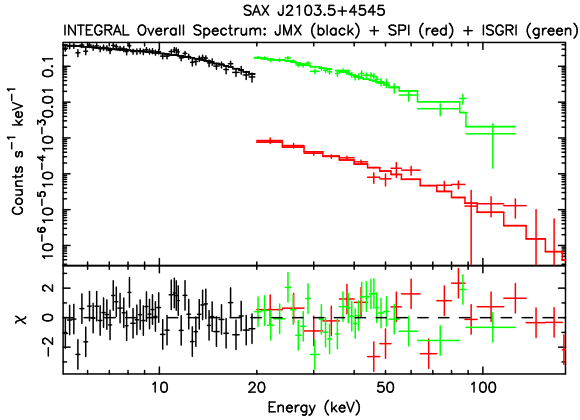}
\caption{SAX~J2103.5+4545 overall spectrum (5--200 keV) 
extracted from JEM-X (5--20 keV), IBIS/ISGRI (20--100 keV, upper spectrum) and SPI (20--200 keV, lower spectrum) instruments (see text). 
The residuals displayed in the lower panel are in units of standard
deviations}
\end{figure}
\end{small}

%%%%%%%%%%%%%%%%%%%%%%%%%%%%%%%%%%%%%%%%%%%%%%%%%%%%%%%

\end{document}